\documentclass[12pt,aps,nofootinbib]{article} 
\usepackage{amsmath,amssymb,physymb}
\usepackage[mathscr]{euscript}
 \usepackage[pdftex]{graphicx}\usepackage{epstopdf}
\newcommand*{\bi}[1]{\boldsymbol{#1}}
\newcommand*{\btau}{\bi{\tau}}
\newcommand*{\bvarsigma}{\bi{\varsigma}}
\newcommand*{\bfeta}{\bi{\eta}}
\newcommand*{\bkappa}{\bi{\kappa}}

\DeclareMathAlphabet{\mathpzc}{OT1}{pzc}{mb}{it}
\newcounter{aaa}

\newenvironment{teor}[2][{}]{\begin{trivlist}\refstepcounter{aaa}%
\labelsep=0pt\item[\bfseries \theaaa. #2. ]#1}
{\end{trivlist}}
\newenvironment{teor*}[2][{}]{\begin{trivlist}%
\labelsep=0pt\item[\bfseries #2. ]#1}
{\end{trivlist}}

\newcommand{\ssy}[5]{#1,    #2 {\bf #3}, #5 (#4)\rlap{.}}
\newcommand{\ogr}[2]{#1\,\vrule \,{}_{\displaystyle{}_{#2}}}
\newenvironment{proof}[1][.
]{\par\noindent\emph{Proof#1}}{\par\nopagebreak
\hfill$\square$\par}

\newcommand{\rmd}{\mathrm{d}}

\newcommand{\Bd}{\mathop{\mathrm{Bd}}}
\newcommand{\Cl}{\mathop{\mathrm {Cl}}}

\newcommand{\vary}{y}
\newcommand{\varg}{g}

\newcommand*{\comp}{{\mathcal K}}
\newcommand*{\compl}{{\mathcal L}}

\newcommand{\ingh}{M^\mathrm{in}}
\newcommand\Z[1]{ \ensuremath{\overset{\looparrowright}{#1}}}                 

\title{Time machines with the compactly determined Cauchy horizon}

\author{S Krasnikov\thanks{Email: S.V.Krasnikov@mail.ru}}%
\date{}
\begin{document}
\maketitle
\begin{abstract}
The building of a  time machine, if possible at all, requires the relevant regions of spacetime to be compact (that is, physically speaking,  free from sources of unpredictability such as infinities and singularities).
Motivated by this argument  we consider the spacetimes with the compactly determined Cauchy horizons (CDCHs), the defining property of which is the compactness of $\overline{J^-(\EuScript U)}\cap J^+(\EuScript S_0)$, where $\EuScript U$ is an open subset of the Cauchy horizon and $\EuScript S_0$ is a Cauchy surface of the initial globally hyperbolic region $\ingh$. The following two facts are established:\\
1)  $\ingh$ has no globally hyperbolic maximal extension. This means that, by shaping appropriately a precompact portion of  a globally hyperbolic region, one can \emph{force} the Universe to produce  either a closed causal curve, or  a quasiregular singularity, whichever it abhors less;\\
2) Before a CDCH is formed a null geodesic appears which infinitely approaches the horizon returning again and again in the same --- arbitrarily small --- region. The energy of the photon moving on such a geodesic  increases with each passage, or at least falls insufficiently fast. As a result,   an observer located in the mentioned   region would see a bunch of photons passing through his laboratory with the  arbitrarily large total energy. We speculate that this phenomenon may have observable consequences.
\end{abstract}
\section{Introduction and discussion}

The quest for a universal mechanism enforcing causality has been  lasting for  a few decades.
The mechanism has never been found and perhaps the time is right  to shift the focus of research from ``why is the time machine impossible?" to ``how to build one?".
A major and widely discussed problem in creating a time machine is its close association with exotic matter. There is, however, a less conspicuous, but seemingly more grave impediment: the only known way to affect the geometry of spacetime (which is to distribute matter so that the Einstein equations would do the job) is ineffectual.
\begin{teor}{Example}\label{DP}
The Deutsch--Politzer (DP) space, see figure~\ref{fig:danger}(a),
\begin{figure}[t,b]\begin{center}
\includegraphics[width= \textwidth]{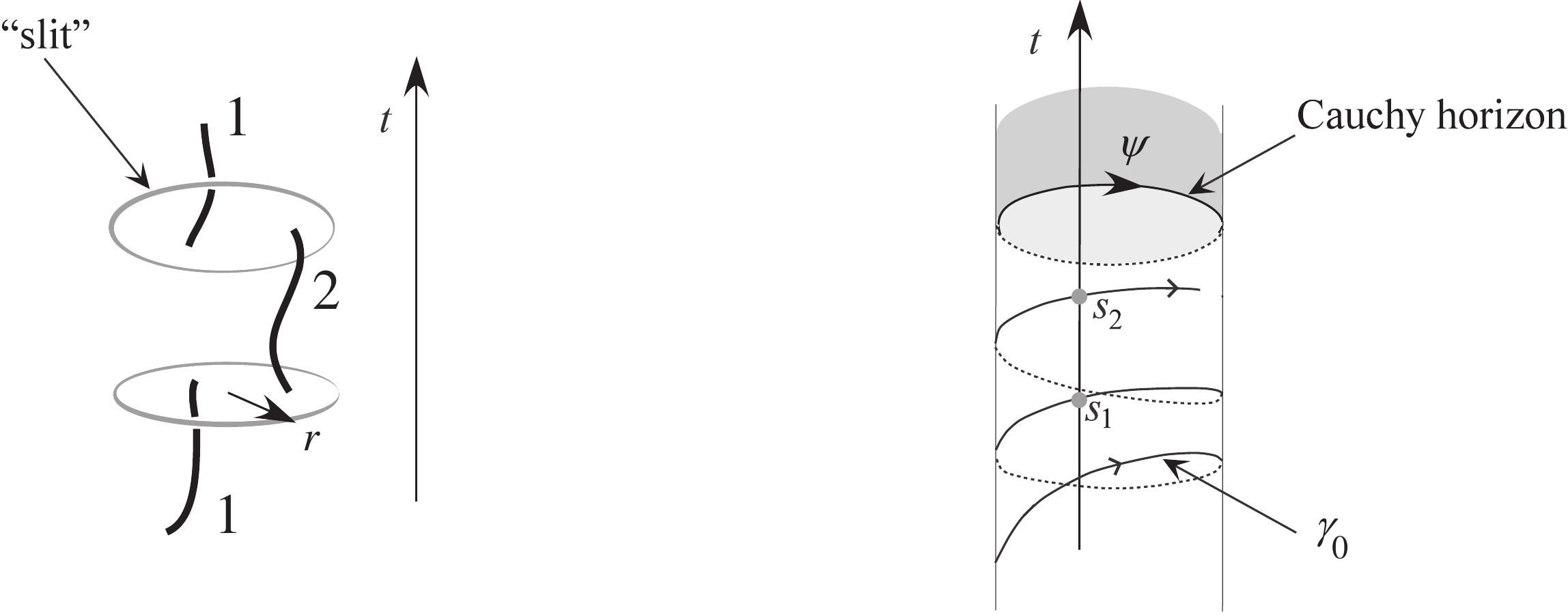}\\
(a)\hspace{0.5\textwidth} (b)
\end{center}\caption{\label{fig:danger} (a) The (2+1)-dimensional Deutsch-Politzer time machine.  Curves 1 and 2 are, actually, continuous and closed, respectively. (b) The Misner time machine.}
\end{figure}
is obtained by, first, cutting  an $n$-dimensional Minkowski space along the unit closed $(n-1)$-disks  at $ t=\pm 1$ and, second, gluing crosswise  the banks of the thus obtained holes:
\[
0\leq r<1,\ t=  1  \pm 0 \quad \leftrightarrow \quad 0\leq r<1,\ t= -1  \mp 0.
\]
 The spacetime is flat and thus its creation does not take any matter (exotic or not) at all. Still no way is seen to build such a time machine. All one can do is ensure that the geometry of the half-space $t<0$
is ``correct" (i.~e., flat). But the decision --- whether  the Universe  will evolve into the time machine or just into a  Minkowski space ---   is not ours.
\end{teor}
Note  the ``slits" in the spacetime: the $(n-2)$-spheres $r=1,\ t=\pm 1$ are missing and cannot be returned back into the spacetime. Their status is much the same as that of closed timelike curves: there are no decisive arguments against them, but according to a certain school of thought, see \cite{quasireg,Ori}, for example, such singularities are unphysical and should be excluded from general relativity (though this has turned out to be unexpectedly hard, see \cite{Ming} for a  concise review).
The null geodesics which constitute the Cauchy horizon, emanate from the slits, which suggests that
  the mentioned unpredictability of the DP space may be  related to their presence,
because ``extra information" comes in from them \cite{Conj}. Be that guess correct, one could simply abandon the DP-like time machines and concentrate on  building a slitless one.
To put it more mathematically let us establish the terminology.
\paragraph{Terminology and conventions}
 Denote by  \Z{M} the set of  all points of a spacetime $M$ at which the causality condition does not hold. Then an inextendible spacetime  $M$ will be called ``time machine"  if
 \begin{equation}\label{eq:def:TM}
\Z{M}\neq\varnothing, \qquad \ingh\equiv  M-\Cl\bigl(I^+(\Z{M})\bigr)  \mbox{ is  globally hyperbolic.}
\end{equation}
$\ingh$ is a past set and its  boundary coincides with its future Cauchy horizon
 \begin{equation}\label{eq:mark}
 \Bd \ingh=\EuScript H^+ (\ingh),
 \end{equation}
 \par\noindent
which  is a closed, imbedded,
achronal three-dimensional $C^{1-}$ submanifold generated by past inextendible null geodesic segments; if  $\alpha_1\neq\alpha_2$ are such segments, then $J^+(\alpha_1)\cap J^-(\alpha_2)$, when nonempty, is their common future end point (these facts --- I drop the proof for the sake of brevity --- follow immediately from the well-known properties of achronal boundaries and Cauchy horizons, see \cite{HawEl} and  \cite{neil}).

 Following \cite{Conj} we call a Cauchy horizon $\EuScript H^+ $  \emph{compactly generated}, if
it has a compact subset
 $\comp\subset
\EuScript H^+$ in which every generator of $\EuScript H^+ $
is totally past imprisoned.
A Cauchy horizon is termed   \emph{compactly determined}, if there is a  Cauchy surface $\EuScript S_0$ of $\ingh$ and an open\footnote{Here ``open" means ``open as a subset of $\EuScript H^+ $"  (i. e., open in the topology induced  on $\EuScript H^+ $ by its embedding in $M$).} set $\EuScript U\subset \EuScript H^+ $ such that the set
$
\compl\equiv \overline{J^-(\EuScript U)}\cap J^+(\EuScript S_0),
$
is compact.
 \par\noindent
Except for examples~\ref{ex:Misner} and \ref{ex:dang} all the spacetimes in this paper are 4-dimensional and the signature of the metrics is $(-,+,+,+)$.

\medskip

It is the spacetime with the    compactly determined \sloppy
  Cauchy horizon (CDCH) that is taken in this paper to be the model of the ``artificial time machine".

Now the conjecture that the unpredictability of the evolution of a spacetime has something to do with the ``slits" can be expressed as a question: ``Is the appearance of a  time machine  with CDCH predetermined by the geometry of its $\ingh$?" It turns out that the answer is still negative. As   proven in \cite{notm}, \emph{the region $\ingh$ of \emph{any} time machine has a causality respecting maximal extension}.\fussy

\begin{teor}{Example}\label{ex:Misner}
Consider  a cylinder $M$ endowed with the (flat) metric
\[
\rmd s^2 =-2\rmd t\rmd\psi-t\rmd\psi{}^2,
 \qquad  t\in \mathbb{R}^1,
 \quad\psi  =\psi+ 1.
 \]
\Z{M} is the half-cylinder $t\geq0$, see figure~\ref{fig:danger}(b),
   and  $\ingh$, called the Misner space in this case, is the globally hyperbolic region $t<0$. So, $M$ is a time machine. Its Cauchy horizon $\EuScript H^+  $ is the circle $t=0$. Choosing   $\comp = \EuScript U$ to be the whole $\EuScript H^+  $ and  $\EuScript S_0$ to be the circle $t=-1$ (which makes $\compl$ be the compact set $-1\leq t\leq 0$), one can easily find that the horizon is both compactly generated and  compactly determined.

  At first glance after the space evolves up to $t=-0$ it cannot help producing a closed causal curve. Consider, however, the spacetime $M'$ obtained by, first, cutting $M$ along  the null ray $\psi=0, \ t\geq 0$, second, cutting a Minkowski plane  along a null ray, and, finally, gluing either bank of the first cut to the corresponding bank of the second. $M'$ is a maximal extension of the Misner space and still   it contains no closed causal curves.
\end{teor}
Thus, no way is seen to ensure that the horizon will prove compactly generated and the spacetime will produce a closed causal curve.
 And yet in a sense  Misner's space \emph{is} more predictable than DP's. One cannot tell what \emph{exactly} information will come to the spacetime and how it will affect the spacetime's evolution, but in the former case we know at least that some extra information \emph{will} come in for sure: the formation of the Cauchy horizon is inevitable, $\ingh$ simply does not have a  globally hyperbolic extension.
\begin{teor}{Definition}\label{def:forved}
 The Cauchy horizon of a time machine
 $M$ is called \emph{forced}, if  $\ingh$ has no globally hyperbolic  maximal   extension. %
\end{teor}
To put it slightly different assume that one prepares a globally hyperbolic spacetime   isometric to the region $\ingh$ of a time machine with the  Cauchy horizon $\EuScript H^+  $. Then, if    $\EuScript H^+  $ is forced, the Universe in its evolution to a maximal extension would have to choose: it would be able to remain  causal or hole-free, but not both. There is no reason to consider the former alternative more probable, see above.
   So, if the procedure proves to create a time machine after all, we could rightfully claim that all the credit for the success is ours. In this sense the time machines with the  forced horizons  are a satisfactory substitute for artificial.
And what makes the compactly generated  Cauchy horizons interesting from the physical point of view is the  fact,  that they  \emph{seem} to be forced. This, however, is yet to be established and that is why we study in this paper another, though quite similar, type of horizon. As will be proven in the next section, all CDCHs \emph{are} forced,

To be compactly determined  a horizon need not be  compactly generated. An example is the spacetime
\[
\rmd s^2 = \rmd^2 y -2\rmd t\rmd\psi-t\rmd\psi{}^2
 \qquad  y,t\in \mathbb{R}^1,
 \quad\psi  =\psi+ 1,
\]
which is a time machine with the horizon  equal to that of the time machine from example~\ref{ex:Misner} multiplied by the real axis. In contrast,
the geometry of a Cauchy horizon which is compactly generated but not   determined is quite bizarre. Even if so pathological spacetimes exist (which is not clear) they definitely do not fit the idea of the ``laboratory-made"  time machine, while this is the only kind of time machines we are interested in. So, there is not much loss of generality in requiring that the Cauchy horizon be compactly determined.

 As we  have seen in Example~\ref{ex:Misner} the fact that a horizon is compactly generated is not a property of $\ingh$, but rather of the \emph{pair}:  $\ingh$ plus its embedding in the extension (indeed,  $\Bd_{M'}\ingh$ has no $\comp$).  In contrast, the fact that a Cauchy horizon  is compactly determined  seems to be independent of the embedding and thus be  an \emph{intrinsic} property of $\ingh$.  Clearly, the intrinsic,  in this sense, properties are of particular interest because they are guaranteed by the geometry of the ``nice" predictable region  $\ingh$. In this paper we study one such property.
 Our concern is a special type of null geodesic which (for the reasons that will become clear in a moment) will be called ``dangerous" and whose defining property is that it:
\begin{enumerate}\renewcommand{\theenumi}{(\emph{\alph{enumi}})}
\item
  lies entirely in   $\ingh$, that is \emph{before} the first closed causal curve appears;
\item returns infinitely many times in $O\cap I^-(p)$, where $p$ is a particular point on  the horizon and $O$ is its arbitrarily small  neighbourhood;
\end{enumerate}
Physically, we interpret $O$, when it is sufficiently small, as a laboratory of the observer, whose world line passes through $p$. If $\gamma_0$ is a dangerous geodesic, the intersection
 $\gamma_0\cap O$ consists of the infinite number of spacelike separated segments  $\tilde\gamma_{i}$ and the observer     will perceive  $\gamma_0$ as a
\emph{bundle} of photons each moving through the laboratory on its own segment  $\tilde\gamma_{i}$.  To assign an energy to such a bundle we pick   a smooth unit timelike vector field $ \btau $ on $M$ (the    field of four-velocities of  fiducial observers) and define the energy of the photon in $\gamma_0( s)$ to be $\varepsilon( s) \equiv -g(\dot\gamma_0( s), \btau( s))$. Assuming
 $O$ to be  small enough to justify neglecting the possible variation of
 $\varepsilon$ along $\tilde\gamma_{i}$ we define  the  ``total energy" of the beam $E\equiv\sum_i \varepsilon(s_{i})$, where $s_{i}$ are picked so that $\gamma(s_{i}) \in \tilde\gamma_{i}$. The value of $E$ depends on the choice of the field $ \btau $, but its finiteness does not. Correspondingly,  the third condition for a geodesic to be called dangerous is
\begin{enumerate}\renewcommand{\theenumi}{(\emph{\alph{enumi}})}\addtocounter{enumi}{2}
\item it returns each time  blue shifted (or at least not too red shifted), so that its $E=\infty$.
\end{enumerate}

 \begin{teor}{Example}\label{ex:dang}
Consider the  null geodesic
\[
\gamma_0:\quad \psi = -2\ln |t| 
 \]
in the Misner space, see example~\ref{ex:Misner}. $\gamma_0$ winds round and round approaching from below the Cauchy horizon. Evidently, for any point $p\in \EuScript H^+$ and any its simply connected neighbourhood $O$  the geodesic is partially imprisoned in  $O\cap I^-(p)=O\cap \ingh$.  For example, it passes through every point of the sequence $\{s_k\colon\ \psi(s_k)=0,\  t(s_k)=-\exp(-k)\}$, $k=1,2,\ldots$. So, $\gamma_0$ possesses the first two properties of dangerous geodesics. Further,
the coordinate transformation
\[
t= -\tfrac{1}{4}{\alpha  \beta  },\qquad \psi=  -2\ln(-\tfrac{1}{2}\beta)
\]
brings the metric and the equation of $\gamma_0$ to the form
\[
\rmd s^2 = -\rmd\alpha \rmd\beta \qquad\mbox{and} \quad
\alpha=-2,
\]
respectively. It is clear now that $\beta$ is an affine parameter on $\gamma_0$ and the
 corresponding tangent vector is $\dot\gamma_0=\partial_\beta=-\tfrac{\alpha}{4} \partial_t-\tfrac{1}{2\beta}\partial_\psi$.
The ``total energy"
$E\equiv\sum_k\varepsilon(s_k)$ is (in)finite when so is
\[
-\sum_kg(\dot\gamma_0, \partial_t)(s_k)= \sum_k\frac{1}{|\beta(s_k)|}.
\]
But the right hand side is equal to $= \frac{1}2\sum_k e^{ k }$, so we conclude that  $ E $ diverges and, correspondingly, $\gamma_0$ is dangerous.
\end{teor}

Another example of a time machine with dangerous geodesics is a simplest version of the Morris--Thorne--Yurtsever time machine \cite{MTY}. On the other hand, their existence in a slightly more general time machine (that with rotating mouths) was disputed in \cite{Nov}. Later, though, it was proven  \cite{Cla} that dangerous geodesics appear in any wormhole-based time machine with the compact $\Z{M}$. In the present paper we generalize considerably the latter result and prove that dangerous geodesics precede \emph{any} CDCH, see propositions~\ref{prop: exist endl} and \ref{prop: cl instab}.
Physically this means that wormholes in the process of their conversion into time machines become sources of  highly collimated high energy beams with the following unusual property.
Though such a beam \emph{looks} as a set of independent photons (in a smoky room one would observe a set of    spacelike separated colorful scintillations),   only one of those photons (that with the highest energy) may end up being absorbed.

 \section{Null geodesics  in the CDCH's past }

\subsection{Doubling of the limit geodesic}

In this subsection we explore the geometry of the immediate past of a CDCH. One of its essential features is the following. From almost all points of the horizon infinitely many past directed null geodesics emanate which are not horizon generators.  All these geodesics must meet the  compact --- in the case of a CDCH --- set $\compl\cap \EuScript S_0$. And this means that we can pick a sequence $\{\gamma_m\}$ of such geodesics which
has two limit curves at once --- a generator of the horizon $\gamma$ and the null geodesic  $\gamma_0$ passing through the limit point of $\{\gamma_m \cap \EuScript S_0\}$. It is the latter, as we shall finally see in the last subsection, that is ``dangerous".
The property of  $\{\gamma_m\}$ to have two limits is inconsistent with global hyperbolicity, and we use this fact to prove that $ \EuScript H^+$ is forced.

Before proceeding further, let us   introduce  a couple  of auxiliary entities. The condition
\begin{equation}\label{eq:sphera}
\varg(\bvarsigma,\btau)=-1,\qquad\quad\mbox{where }
\bvarsigma\equiv \partial_ l,
\end{equation}
fixes   (up to an additive constant) the ``arc length parameter''
$ l $ on any smooth causal curve $\mu( l)$. The name is due to the fact that
 \[
d_{ l _1 l _2}<| l _1- l_2|< \sqrt2d_{ l _1 l _2},
\]
 where $d_{ l _1 l _2}$ is the length of the segment $\evalat{\mu}{l\in[ l _1,l_2]}{}$ in the Riemannian metric
\begin{equation}\label{eq:auxRiem}
\varg^R(\bi
x,\bi y)\equiv \varg(\bi x, \bi y) + 2\varg(\bi x, \btau)\varg(\btau,\bi y),
\end{equation}

\begin{teor}{Proposition} \label{prop: exist endl}
In any time machine with the compactly determined Cauchy horizon there exists a future inextendible null geodesic \sloppy
$\gamma_0$, totally imprisoned in the set $\compl\cap\ingh$.
\end{teor}

\begin{proof} \fussy
 Denote by $\mathsf H$ the set of all points  which
have the property
that all past directed null geodesics  emanating from them
 lie in $\EuScript H^+$:
\begin{equation}\label{eq:set P}
\mathsf H\equiv\{x\in\EuScript H^+\colon\quad \bigl(J^-(x)- I^-(x)\bigr)\subset  \EuScript H^+\}.
\end{equation}
Note that
\begin{equation}\label{eq: H isol}
 \text{ $\mathsf H$, when non-empty, consists of isolated points,}
\end{equation}
since for any $x\in \mathsf H$ the set  $J^-(x)- I^-(x)$ is a neighbourhood of $x$ in $\EuScript H^+$ which does not contain other points of $\mathsf H$ [indeed, if $y\in \bigl(J^-(x)- I^-(x)\bigr)$, then according to the relevant property of the Cauchy horizons, see below eq.~\eqref{eq:mark}, the only generator of the horizon passing through $y$ is the extension of the null
    geodesic segment from $x$ to $y$. All other null geodesics terminating
    at  $y$ do not lie in  $\EuScript H^+$ and, correspondingly, $y\notin \mathsf H$].

Pick a point $p\in (\EuScript U-\mathsf H)$. The velocities in $p$ of the horizon generators   parametrized by $l$ form an --- obviously closed --- proper subset of the sphere \eqref{eq:sphera}. Hence there are null future directed geodesics $\gamma$ and  $\{\gamma_m\}$, $m=1,2\ldots$ such that they all terminate in $p$ and
\[
\gamma\subset \EuScript H^+,\qquad
\dot \gamma_m(p)  \to \dot \gamma(p),\qquad \forall m\quad (\gamma_m-p)\subset\ingh.
 \]
Every $\gamma_m$  meets $\EuScript S_0$ in some point $q_m$ and we shall use the latter as the origin of the  length parameter on $\gamma_m$  (from now on $\gamma_m$ is understood to be the segment from $\EuScript S_0$ to $p$, rather than the whole geodesic):
 \[
 l\in [0, l_m^{\max}],\qquad \gamma_m(0)= q_m, \quad
 \gamma_m( l_m^{\max})= p.
 \]
 All $q_m$ lie in the compact (by the definition of the CDCH) set
$\overline{J^-(\EuScript U)}\cap\EuScript S_0=\compl\cap \EuScript S_0$, see figure~\ref{fig:M}a,
\begin{figure}[t,b]\begin{center}
\includegraphics[width=0.9\textwidth]{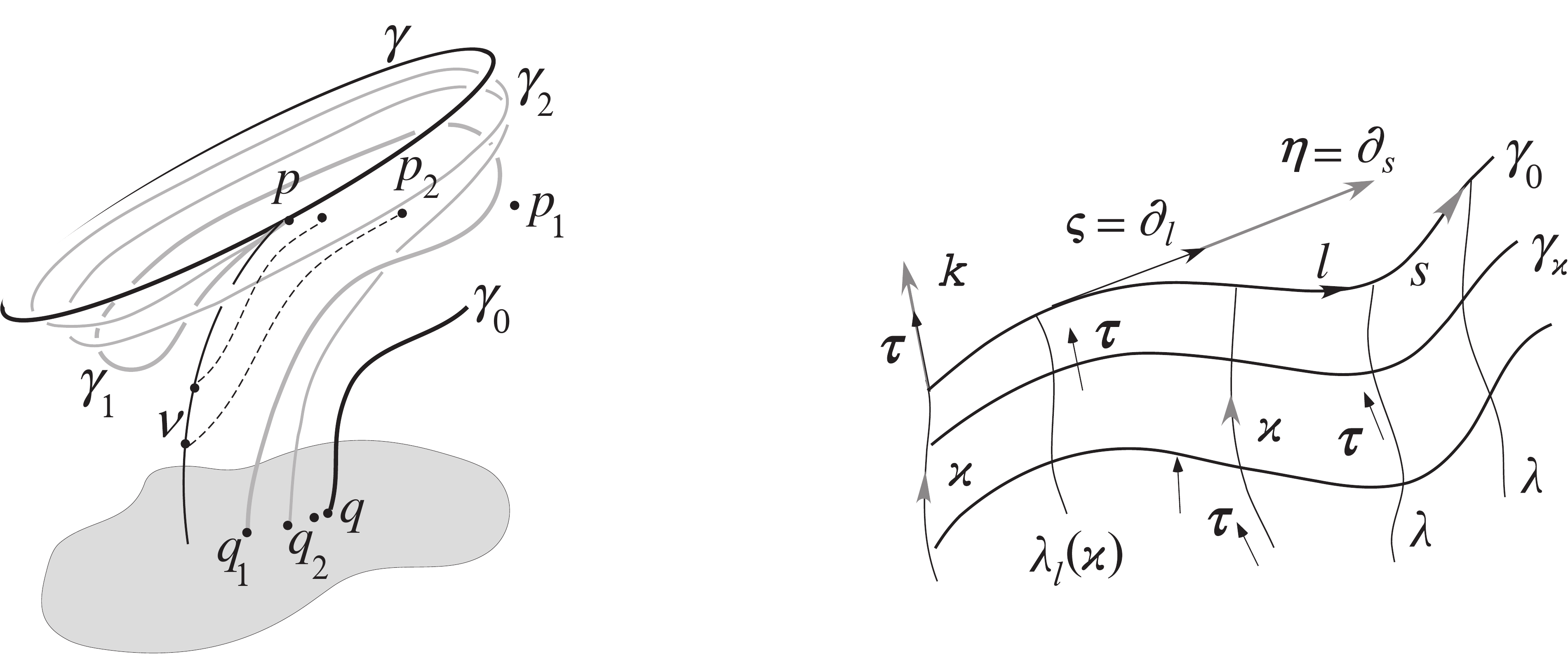}\\
(a)\hspace{0.5\textwidth} (b)
\end{center}\caption{\label{fig:M} (a)  The shadowed spot is the set $\compl\cap \EuScript S_0$. The dashed lines are $\chi_i$. (b) The homotopy $\Lambda$. The curve $\gamma_0$ is null and $\lambda$s are timelike.}
\end{figure}
so passing to a subsequence, if necessary, we state that
\[
q_m\to q, \qquad
\partial_ l (q_m)\to {\bvarsigma},
\]
where $ q$ --- is a point of $\compl\cap \EuScript S_0$, and $ {\bvarsigma}$
 is a null vector satisfying \eqref{eq:sphera}.
We define $\gamma_0( l )$ to be the future inextendible
geodesic fixed by the conditions
\[
\gamma_0(0)=q,\qquad \dot\gamma_0(q)= {\bvarsigma}
\]
and our task now is to prove  that it is, indeed, totally imprisoned in
$\compl\cap\ingh$.

At any $ l<\varlimsup   l_m^{\max}$ there is a sequence of points $\gamma_m( l )$ which  by the compactness of $\compl$  contains a convergent subsequence $\{\gamma_j( l )\}$. Its limit is $\gamma_0( l)$ since $\gamma_0 $ is a solution of the geodesic equations and their solutions are known to depend continuously on the initial conditions,
So, $\gamma_0 $
cannot leave the set $\Gamma\equiv\overline{\strut\cup_j \gamma_j}$ as long as $l\leq
\varlimsup  l_j^{\max}$,
that is until it passes through
$\gamma_0(\varlimsup l_j^\text{max})
=p$. But the latter is impossible,
because  it would imply  the equality $\gamma_0=\gamma$ (since these two geodesics would emanate with the same velocities  $\lim\dot\gamma_m$ from the common point   $p$) in  contradiction to the fact that
$\gamma$ being a generator does not leave $\EuScript H^+$ in the past direction, see the Introduction. Hence
$
\gamma_0\subset \Gamma\subset
\compl,
$
and it remains only to prove that $\gamma_0$  does not leave $\ingh$ either.

To derive a contradiction assume that $\gamma_0$ does leave $\ingh$ and, correspondingly, meets $\EuScript H^+$ in a point $r=\gamma_0( l_r)$, where $ l_r<\varlimsup l_j^\text{max}$. Then what will happen with $\gamma_0$ to the future of  $r$? Three possibility are conceivable:
(\emph{a}) $\gamma_0$  enters $M-\overline{\ingh}$; (\emph{b}) it remains in $\EuScript H^+$; (\emph{c})
it returns to $\ingh$.
But in the case (\emph{a})   some  $\gamma_j$ would also enter $M-\overline{\ingh}$, which is impossible, since $\gamma_j(l_j^{\max}-0)\in\ingh$ and $\ingh$ is a past set. The possibility (\emph{b}) is excluded,  too, otherwise  $\gamma_0$ would be a (part of a) horizon generator to the future of $r$, so it would not be able to leave the horizon in the past direction. Finally, (\emph{c}) is also impossible:
moving to the past along a causal curve ($\gamma_0$  in this case) from a point of the open past set $\ingh$,  one can not reach  the boundary of the latter.
\end{proof}

\begin{teor}{Proposition}\label{pr: lab}
Any compactly determined Cauchy horizon is forced.
\end{teor}
\begin{proof}
To obtain a contradiction assume that there is an isometry $\varpi$.
 mapping $\ingh$ to a proper subset ${\hat M}$ of some globally hyperbolic inextendible spacetime $ M^e$. From now on to simplify notation we write $\hat A$ for $\varpi(A)$, where $A$ is any subset of $\ingh$.
 Note that the point $p^e$ and the geodesic $\gamma^e$ defined below  lie off ${\hat M} $ and that is why they are not denoted $\hat p$ and $\hat \gamma$, respectively.

Let us  start with the observation that  any neighbourhood $ U $ intersecting   $\EuScript H^+$ must contain a sequence   $\{p_k\}$, such that
\begin{equation}\label{eq:seq}
 p_k\to   p\in  \EuScript H^+ ,
\qquad
\hat p_k\to   p^e\in  \Bd{{\hat M} }, 
\end{equation}
because otherwise we could, contrary to our hypothesis, extend the spacetime $M^e$ by gluing $U$ to it with the isometry $\varpi$. Indeed, the space $U\cup_\varpi M^e$ is obviously a smooth connected Lorentzian manifold. So, it is a spacetime --- and hence an extension of $M^e$ --- if it is Hausdorff. And this latter property does hold  unless there exists a sequence \eqref{eq:seq}, see \cite[proposition III~1.6]{Alex}. Further, $U$ has been chosen arbitrary, whence it follows  that the set of points $p$ satisfying \eqref{eq:seq} is dense in $\EuScript H^+$. Which means, when the property \eqref{eq: H isol} is taken into account, that there is a point among them which lies in $\EuScript U - \mathsf H$. It is this point that will be denoted by $p$ in what follows, while $q$, $q_m$, $\gamma$, and
$\gamma_m$,  $m=1,2\ldots$  are the same points and geodesic segments as in the proof of proposition~\ref{prop: exist endl}, we only change now the definition of the segments  $\gamma_m$ by excluding $p$ from them.

Now consider a future directed timelike curve $\nu(\upsilon)$ terminating at $p$. Let us prove that the future end point of $\hat \nu$ is $p^e$. For this purpose, let us introduce one more family of geodesics. Pick a sequence of points $\nu(\upsilon_i)$ converging to $p$ and connect each of them to a point $p_{k(i)}$ by a geodesic  $\chi_i(\xi^{(i)})$. The numbers $k(i)$, curves $\chi_i$,  and their affine parameters $\xi^{(i)}$ are chosen so that beginning from some $i_0$:
\begin{subequations}
\begin{gather}
\label{eq:b_timeli} \chi_i   \quad\text{are timelike and lie in $O$};
\\
\label{eq:aff_norm}
 \varg(\bi z,\partial_{\xi^{(i)}})=-1 \quad\forall \upsilon_i,
\end{gather}
\end{subequations}
 where $O$ is a normal neighbourhood of $p$ and $\bi z$ is some timelike vector parallel translated along  $\nu$. Condition \eqref{eq:b_timeli} combined with the fact that $\ingh$ is a past set ensures the inclusion $\chi_i \subset\ingh $ and, as a consequence, the existence of $\hat \chi_i(\xi^{(i)})$ for all $i$. Denote the affine lengths of $\chi_i$ and $\hat \chi_i$ (i.~e., $\xi^{(i)}[\overset{
 \mbox{ \tiny$\scriptscriptstyle(\wedge)$}
 }{p}_{k(i)}] - \xi^{(i)}[\overset{ \mbox{ \tiny$\scriptscriptstyle(\wedge)$                         }
 }{\nu}(\upsilon_i)]$) by $L( \chi_i)$ and  $L( \hat \chi_i)$, correspondingly. Evidently, $L(\hat \chi_i)=L(\chi_i)$ and $L( \chi_i)\to 0$, which implies
\begin{equation}
\label{eq:Llim}
L(\hat \chi_i)\to 0\qquad\text{at }i\to\infty.
\end{equation}
Now note that for any $i_0$ all $\hat p_j$ with  $j> k(i_0)$ are in the chronological future of
 $\hat\nu(\upsilon_{i_0})$. So,
\[
\forall\upsilon\qquad p^e\in\overline{I^+(\hat\nu(\upsilon))} =\overline{ J^+(\hat\nu(\upsilon))} = J^+(\hat\nu(\upsilon)) \text{ and hence }
\hat\nu(\upsilon) \in J^-(p^e),
\]
where the last equality follows from the global  hyperbolicity of $M^e$.

Thus, $\evalat{\hat\nu}{\upsilon>\upsilon_0}{}$ is imprisoned in the compact set $J^-(p^e)\cap J^+(\hat\nu(\upsilon_0))$. Which means that the sequence  $\hat \nu(\upsilon_i)$ (or a  subsequence of it)  has a limit and this limit must coincide with  $p^e$ being connected with the latter by a geodesic with the finite [as follows from \eqref{eq:aff_norm}] initial velocity and zero [as is seen from \eqref{eq:Llim}] parameter length. The same reasoning applies to the case when $\{\upsilon_i\}$ is replaced by any other increasing sequence tending to the same limit\footnote{As the example of the Misner time machine shows,   for a \emph{non} globally hyperbolic $M^e$ even that is not necessarily the case.}. So, $p^e$, indeed, is the future end point of  $\hat \nu$. Moreover, the proof remains valid with $\{p_k\}$ replaced by $\{\gamma_m(l_k)\}$ with $l_k\to  l_m^\text{max}$ and $p^e$ defined to be the end point of $\hat \nu$. From this results   the conclusion which completes the first part of the proof:
\emph{$p^e$ is the common future end point of all $\{\hat\gamma_m\}$}.

Now let us examine the boundedness of the set of the affine lengths of the geodesics $\{ \gamma_m\}$ and $\{\hat\gamma_m\}$. To this end fix the affine parameter $s^{(m)}$ or $\hat s^{(m)}$, respectively, on each of those geodesics by the conditions
 \[
s^{(m)}(p)=\hat s^{(m)}(p^{e})=0, \qquad \varg(\partial_\upsilon, \partial_{s^{(m)}})(p) = \varg(\partial_\upsilon, \partial_{\hat s^{(m)}})(p^{e}) =1
 \]
(the second condition means, in particular, that $ \gamma_m $ are past directed) and denote for brevity  $ L_m\equiv s^{(m)}(q_m)$, $\hat L_m\equiv s^{(m)}(\hat q_m)$. The sequence $\{ L_m\}$ is unbounded, because otherwise $\gamma $ would pass through $q\in\EuScript S_0$ at some finite positive $s$, which is impossible,  since  $\gamma$ cannot leave $\EuScript H^+$ being its generator. But $L_m=\hat L_m$ and hence the mentioned  unboundedness implies the unboundedness of $\{ \hat L_m\}$. Which results in the fact that the geodesic
\[
\gamma^e( \hat s)\equiv\lim_{\substack{m\to\infty\\  \hat s^{(m)}\to  \hat s}} \hat\gamma_m ( \hat s^{(m)})
\]
will never meet $\varpi(\EuScript S_0)$ and, as a consequence, never leave the compact set  $J^-(p^e)\cap J^+\bigl[\varpi (\compl\cap \EuScript S_0 )\bigr]$. And according to \cite[proposition 6.4.7]{HawEl}  this contradicts the strong causality of  $M^e$ (which follows from the global hyperbolicity of the latter).

\end{proof}

\subsection{Deformation of imprisoned geodesics}
 Let us consider the curves  obtained by moving every point of $\gamma_0$ some distance to the past along the corresponding
integral curves of $\btau$.
Evidently, this deformation can be done so that the resulting curve ---   denote it  $\gamma_*$ --- remains nonspacelike (it suffices, roughly speaking, to require  the mentioned distance  to decrease with $l$) and  imprisoned in a compact subset of $J^-( \compl )$. At the same time $\gamma_*$  cannot get imprisoned in any compact  subset of
   $I^-( \compl )$, see~\cite[proposition 6.4.7]{HawEl}.
    The proof of   proposition~\ref{prop: cl instab} will rest on this ``contradiction" the main step being the proof of a lemma  which  enables us to find the  mentioned deformation explicitly. To realize this program we  need a number of new objects.

 On $\gamma_0$ in addition to  $ l $
define an \emph{affine} parameter $ s$ (the definition of  $s^{(m)}$ cannot be extended to the case $m=0$)   so that the velocity
$\bfeta\equiv\partial_ s$  is future directed and $ s=0$   at  $q$. Now $\gamma_0$  is characterized by the (evidently negative) function
\[
h\equiv  \eta^a\tau_a,
\]
which relates $ l $ to $ s$:\fussy
\begin{equation}\label{def:h}
h=-\frac{\rmd  l }{\rmd s},\qquad  s( l )=\int_ l
^0\frac{\rmd  \breve l }{h(\breve l )}
\end{equation}
and one velocity to the other:
\begin{equation}\label{eq:l s vars}
    \bfeta = -h\bvarsigma.
\end{equation}
The function  has a transparent physical meaning (unless $\gamma_0$ is a loop):
\begin{equation}\label{eq: enerPhot}
h( l _1)/h( l _2)=
\varepsilon( l _1)/\varepsilon( l _2),
\end{equation}
where $\varepsilon( l )$  is the energy at a point $\gamma_0( l )$ of a photon moving on $\gamma_0$, as measured by an
observer with the velocity $\btau[\gamma_0( l )]$.

Pick a positive (and sufficiently small, see below) constant $\varkappa_*$ and a   smooth  function $f$ defined on the non-negative semiaxis $\mathbb R_+$ and obeying the inequalities $\underline f \leq f
\leq  \overline f $,  where   $\underline f $, $\overline f \in \mathbb R_+$ are  constants. Define a homotopy
\[\Lambda( l ,\varkappa)\colon\quad\mathcal G\to M,\qquad \mbox{ where }\mathcal G\equiv \mathbb R_+\times[-\varkappa_*,0],
\]
see figure \ref{fig:M}(b), by requiring that\label{p:Lambda}
\begin{enumerate}
\parskip=1em
\renewcommand{\labelenumi}{(\emph{\alph{enumi}})}
\renewcommand{\theenumi}{(\emph{\alph{enumi}})}
\labelsep=0.5em
  \item the first  ``horizontal'' curve be  $\gamma_0$:\label{usl:1hor}
      \[\Lambda( l ,0)=\gamma_0( l );\]
  \item each ``vertical'' curve
      $\lambda_c(\varkappa)\equiv \Lambda(c,\varkappa)$ be
     (a part of) an integral  curve of the field~$\btau$\label{usl:eta-coord};
  \item
  the velocity
      $\bi\kappa=\partial_\varkappa $ in every point $p\equiv\lambda_ l
      (\varkappa)$ be equal to $f( l )\btau(p)$.\label{usl:param k}
\end{enumerate}
In other words,  $\Lambda$ is constructed so that for any $\varkappa $ the ``horizontal''  line $\gamma_{\varkappa }( l )\equiv \Lambda( l
,{\varkappa })$   is obtained from $\gamma_0$ by moving each its point  to the past along the integral curves of $\btau$ by the distance $f( l )|\varkappa |$ (recall that all $\varkappa$ are negative) in  the natural parameter (loosely speaking $f$ defines the shape of the deformation and $\varkappa$ --- its amplitude).

Thus, we have introduced two types of curves:
   \begin{enumerate}
            \item  Horizontal  curves $\gamma $. One of them---specifically, $\gamma_0$---is parameterized by  the ``arc length parameter''  $ l$ defined by
  \eqref{eq:sphera}. This curve (perhaps, in contrast to the other $\gamma $s) is a null
  geodesic. In addition to
  $ l$ there is an \emph{affine} parameter
 $ s$ on $\gamma_0$. All  other  horizontal curves
  are parameterized only by $ l$, which in this case is defined by the requirement
[see item \ref{usl:eta-coord} in the definition of $\Lambda $] that it is constant along  $\lambda $'s (note that when
$\varkappa $ is non-zero, $ l$ need not be an arc length parameter on $\gamma_{\varkappa}$). The velocity vector corresponding to
 $ l$ is denoted by~$ \bvarsigma$;
            \item  Vertical curves
  $\lambda $ which are the integral curves of the field $\btau$.
  In addition to the natural parameter there is also  the parameter
  $\varkappa$ on them which differs from the former only by a constant (on each  $\lambda $) factor, (see item~\ref{usl:param k} in the definition of $\Lambda $).
  The velocity vector corresponding to $\varkappa$
  is denoted by~$\bkappa$;
          \end{enumerate}
  and a scalar function $f$ which is initially defined  on  $\mathbb R_+$, but which we extend now to the entire $\mathcal G$ by the relation $f(p)\equiv f[l(p)]\quad\forall p\in \mathcal G$.

\begin{teor}{Technical notes} 1) Let
$t_\mathrm{m}(p)$ be the length, with respect to the Riemannian metric $\varg^R$, of the shortest inextendible curve emanating from $p$. This quantity, if finite, is continuous and positive in the whole $\compl$. Then the compactness of $\compl$ implies  $\inf_{\compl}t_\mathrm{m}> 0 $. In other words,  from any point of $\compl$ it is possible to travel any distance $T<\inf_{\compl}t_\mathrm{m}$  (in natural parameter)  along the integral curve $\lambda$ of the field   $\btau$.
Thus the existence   of the homotopy $\Lambda$ is guaranteed for any  $ \varkappa_*< \inf_{\compl}t_\mathrm{m}/ (2\overline f)$.\\
2)
The surface $\Lambda(\mathcal G)$ may have self-intersections, so it should be noted that,
first, we consider $\bfeta$ and $h$ as functions of $ l$, not of a point of $M$.
And, second, the vectors $\bvarsigma  $ and $\bkappa $ are maps $ \mathcal G  \to T_M$. Correspondingly,  a derivative like ${\varsigma ^a}_{;b}(p)$, $p\in  \mathcal G $ is actually shorthand for
$
\bigl[\varsigma ^a \circ \tilde\Lambda^{-1} \bigr]_{;b}\bigl(\Lambda (p)\bigr)$,
where $\tilde\Lambda$ is the restriction of $\Lambda$ to a neighbourhood of $p$ in which $\Lambda$ is injective.
\end{teor}

Our way of defining  $l$ gives rise to a useful relation. To derive it pick a coordinate system $\{l, \varkappa, x_1,x_2\}$ in a convex neighbourhood of $p$  so that $\Lambda(\mathcal G)$ be the surface $x_{1,2}=0$.
Then the    curves $\lambda $ and
$\gamma $ are coordinate lines and hence their tangent vectors commute:
\begin{subequations}
\begin{equation}
 {\varsigma ^a}_{;\varkappa}- {\kappa^a}_{; l }\equiv
{\varsigma ^a}_{;b}\kappa^b-{\kappa^a}_{;b}\varsigma ^b
 = {\varsigma ^a},_{\varkappa}- {\kappa^a},_{ l }=
0,
\label{ur10}
\end{equation}
which gives in particular,
\begin{equation}\label{ur9}
{\varsigma _a}_{;\varkappa}\equiv\varsigma
_{a;b}\kappa^b=\kappa_{a;b}\varsigma ^b=(f\tau_{a})_{;b}\varsigma ^b
=f' \tau_{a} + f\tau_{a;b}\varsigma ^b.
\end{equation}
\end{subequations}

In proving  proposition~\ref{prop: cl instab}, we shall  need a compact set containing $\gamma_0$ to lie in $\ingh$, not just in  $\compl$ (which contains some points of the horizon). To  formulate a convenient sufficient condition for the existence of such a set consider a map $F$ which sends every pair
\[
 \lambda_l (\tau)\in\compl
, \qquad t\in [ \overline f \varkappa , \underline f \varkappa]
\]
to the point $\lambda_l(\tau + t)$.
      Any $\gamma_\varkappa$ with $\varkappa\in [-\varkappa_*,0]$ lies in   ${\mathcal O}\equiv F\bigl( \compl\times
[  \overline f \varkappa ,\underline f \varkappa]\bigr)$.
  Being the image of a compact set under a continuous map, $\mathcal O$ is compact. Also it lies, by construction, in $J^-( \compl)$. Generally,  it may happen, that $\mathcal O\not \subset I^-( \compl)$, but such a possibility is excluded, if
 $F$ shifts \emph{all} points of $\compl$. Thus we have established the existence of a  compact set $\mathcal O$ such that
\begin{equation}
\label{eq:Lam v comp}
\begin{split}
 &\forall\varkappa\in [-\varkappa_*,0]\quad\qquad
\mathcal O\supset\gamma_\varkappa ,
\\
&\mbox{If }\underline f , \varkappa\neq 0,\qquad \mbox{then\ } \mathcal O\subset I^-( \compl).
\end{split}
\end{equation}

\begin{teor}{Lemma} \label{lem: part of 8.5.5}
If $f'/f$  is bounded and for some positive constant
$c_1$
\begin{equation}\label{eq:ogr na f}
h'/h< - f'/f -c_1  f
 ,\qquad
\forall   l \in \mathbb R_+,
\end{equation}
then there is $\varkappa_0$ such that the curve $\gamma_{\varkappa_0}$   is timelike and future inextendible.
\end{teor}
Before proceeding to the proof proper  (it will be similar to the proof of  \cite[lemma~8.5.5]{HawEl}) we have to establish the boundedness --- at sufficiently small $\varkappa_*$ --- of
a number of relevant quantities. First, note that $f$ and $f'$ are bounded, by definition and
by hypothesis, respectively.
And so are $\tau^a$, ${\tau^a}_{;b}$ and
${\tau^a}_{;bc}$, since they are smooth in the entire compact
set $\mathcal O$, which, as follows from \eqref{eq:Lam v comp}, contains $\Lambda({\mathcal G})$ (for the sake of simplicity in discussing boundedness
of tensor components we shall assume that the whole $\mathcal O$ is covered by a single coordinate system. The generalization to the case when   a few such systems are necessary  is
straightforward, and owing to the compactness of $\mathcal O$  one never needs infinitely many charts).
Next, consider the components
$\varsigma ^a$. At $\varkappa=0$ the boundedness of $\varsigma^a
$ or, equivalently, of  $\varsigma ^a_\perp \equiv \varsigma
^a+\tau^a$ follows from
the fact that the length of $\bvarsigma _\perp$
in the Riemannian metric~\eqref{eq:auxRiem}  is constant on $\gamma_0$:
\begin{multline*}
\varg^R(\bvarsigma _\perp,\bvarsigma _\perp)=\varg(\bvarsigma _\perp,\bvarsigma _\perp) + 2[\varg(\bvarsigma
_\perp,\btau)]^2= \varg(\bvarsigma ,\bvarsigma ) + \varg(\btau,\btau)  + 2\varg(\bvarsigma ,\btau) +
\\
  + 2[\varg(\bvarsigma ,\btau)+ \varg(\btau,\btau)]^2=5.
\end{multline*}
Next, the function $\varsigma ^a( l ,\varkappa)$  is bounded also on the entire
  $\mathcal G$, being the solution of the differential equation ~\eqref{ur9} with the bounded coefficients and with the bounded, as we have just established,  initial value
$\varsigma ^a( l ,0)$.
Finally, the chain
\begin{equation}\label{eq: ur na h}
h'=( \eta^a\tau_a)_{;b}\evalat{\varsigma ^b}{\gamma_0}{}=
\eta^a\tau_{a;b}\varsigma ^b=
-h\varsigma ^a\tau_{a;b}\varsigma ^b,
\end{equation}
in which the second equality follows from the fact that $\gamma_0$ is a geodesic, and the last one --- from \eqref{eq:l s vars}, proves that \emph{irrespective of the validity of \eqref{eq:ogr na f}}
\begin{equation}\label{eq: h'/h bound}
  h'/h\quad\mbox{is bounded on } \gamma_0.
\end{equation}

\begin{proof}[ of the lemma.]
Consider the  function $\vary \equiv \varsigma ^a\varsigma _a$. It plays the role of an indicator:
a curve $\gamma_{\varkappa}$ is timelike at some
 $l_0$ when and only when $\vary( l_0
 ,\varkappa)$  is negative.
Write down the following chain of equalities valid in all
points of $\gamma_0$
\begin{multline}
 {\textstyle \frac12} \vary ,_\varkappa=\varsigma _{a;\varkappa}\varsigma ^a
=\kappa_{a; l }\varsigma ^a= (f\tau_a)_{; l }\ogr{\varsigma
^a}{\gamma_0}{}=-f' -f\tau_a{\varsigma ^a}_{; l }
=-f'+f\tau_a(h^{-1} \eta^a)_{; l }
\\=
 -f' + fh(h^{-1}),_ l +f\tau_ah^{-1}{
\eta^a}_{; l } =-f' - f\ln'|h|,\label{eq: ll,a}
\end{multline}
which is derived by using, in turn, equation~\eqref{ur10}
(in the second equality), the normalizing condition \eqref{eq:sphera} defining $ l$ (in the penultimate equality of the first line),
the relation \eqref{eq:l s vars}  (in  the next equality)  and, finally,
the fact that $\bfeta$ satisfies the geodesic equation
${ \eta^a}_{; l }=-\frac1h{ \eta^a}_{;b} \eta^b=0$ on $\gamma_0$. Combining the resulting equation with the hypothesis~\eqref{eq:ogr na f} one gets
     \begin{equation}\label{eq:y'>0}
\vary,_{\varkappa}( l,0) > 2c_1f^2.
\end{equation}
On the other hand, $\vary =0$  on $\gamma_0$.
So, we conclude that for any $ l $ there is (negative, of course) $\varkappa_{**}$
--- its value depends on $ l $ --- such that
\begin{equation}\label{eq:timel-ss}
\vary ( l ,\varkappa)<0,\qquad\forall
\varkappa\in(\varkappa_{**}( l ),0).
\end{equation}
Thus, $\gamma_\varkappa$ for the relevant
$\varkappa$ is timelike at $ l $.  We, however, are looking
for an \emph{inextendible} timelike curve or, equivalently, for
a $\varkappa_{**}$ such that
the inequality~\eqref{eq:timel-ss} holds at all $ l $
\emph{simultaneously}. So, let us write down one chain of equalities more
[the last parenthesized factor is ${\varsigma ^a}_{;\varkappa}$
transformed with the use of \eqref{ur9}]:
\begin{eqnarray*}
\forall\,\varkappa\qquad {\textstyle \frac12}\vary ,_{\varkappa\varkappa} =
[(f\tau_a)_{; l }\varsigma ^a],_\varkappa =
 (f\tau_a)_{; l \varkappa}\varsigma ^a +(f\tau_a)_{; l }\varsigma ^a_{\;\;;\varkappa}
\\
=f'\tau_{a;\varkappa}\varsigma ^a +f\tau_{a; l \varkappa}\varsigma ^a
 +(f'\tau_a+ f\tau_{a; l })( f'\tau^a +  f{\tau^a}_{;b}\varsigma ^b).
\end{eqnarray*}
Substitute the formulas 
\begin{eqnarray*}
\tau_{a; l }=\tau_{a;b}\varsigma ^b, \qquad
\tau_{a;\varkappa}=\tau_{a;b}\kappa^b=f\tau_{a;b}\tau^b,
\\
\tau_{a; l \varkappa}=\tau_{a;bc}\varsigma ^bf\tau^c +
\tau_{a;b}{\varsigma ^b}_{;\varkappa}=f\tau_{a;bc}\tau^c\varsigma ^b
+ \tau_{a;b}(f'\tau^{b} + f{\tau^b}_{;c}\varsigma ^c),
\end{eqnarray*}
in the rightmost part of the chain to obtain
\begin{equation*}
\vary ,_{\varkappa\varkappa}( l ,\varkappa)= b_1 f'^2 +
b_2 f'f + b_3f^2,
\end{equation*}
where $b_k$ are some bounded (as follows from the boundedness of  $\varsigma
^a$  and the derivatives of $\tau^a$, established above) functions.
Dividing this by the inequality \eqref{eq:y'>0}  we get (recall that by hypothesis $f'/f$ is bounded)
\begin{equation}\label{eq:ravnom}
|\vary ,_{\varkappa\varkappa}( l ,\varkappa)/\vary
,_{\varkappa}( l,0)|<c_2,
\quad\qquad\forall\varkappa\in [\varkappa_{**},0],
\end{equation}
where  $c_2$ is a constant. Since $\vary (l,0)=0$ and $\vary
,_{\varkappa}(l,0)>0$, as we already know,
   \eqref{eq:ravnom} means that for some
negative  $\varkappa_0$
\[
 \vary ( l,\varkappa_0)<0,\qquad\forall  l \in\mathbb R_+,\ \varkappa\in[\varkappa_0,0),
\]
and hence $\gamma_{\varkappa_0}$ is \emph{timelike}.

Similar arguments apply to the quantity $\omega \equiv
\varsigma ^a\tau_a$. Namely, $\omega =-1$ at $\varkappa=0$, while its derivative in
the $\bkappa$-direction
\[
\omega_{;\varkappa}= {\varsigma}_{a;\varkappa}\tau^a+
\varsigma^a\tau_{a;\varkappa}=
f'\tau_{a}\tau^a + f\tau_{a;b}\varsigma ^b\tau^a+
f\varsigma^a\tau_{a;b}\tau ^b
\]
is bounded. Hence, $\omega ( l ,\varkappa_0)$ at sufficiently small $\varkappa_0$ is greater (in absolute value) than
$\frac12$. Consequently, the length of the corresponding $\gamma_{\varkappa_0}$ in the Riemannian  metric \eqref{eq:auxRiem} is infinite, which evidently means that
$\gamma_{\varkappa_0}$ is \emph{inextendible}.
%
\end{proof}
\subsection{The danger of $\gamma_0$.}
In this subsection we use  the just proven lemma to demonstrate how pathological the energy
properties are of the photon traveling on $\gamma_0 $.  Assign the unit value to the energy of the photon in a point $\gamma_0(l_0)$ (the energy thus becomes a function of two variables ---
$ l$ and $l_0$). Then split  $\gamma_0 (l)$ into segments of   unit parameter length, using \eqref{eq: enerPhot} find the maximal   energy on each segment  and sum those maximal values up.
It turns out that for an arbitrary constant $E$ an appropriate choice of the initial point $\gamma_0(l_0)$ will make the sum (that is  the  ``total energy"   discussed in the Introduction) greater than $E$.
\begin{teor}{Proposition} \label{prop: cl instab}
Let $\gamma_0( l )$ be as in  proposition~\ref{prop:
exist endl}. Then for an arbitrarily big constant $E$ there is a positive number $l_0$ such that
\begin{equation*}
\frac{1}{|h(l_0 )|}\sum_{k=0}^\infty h_k \geq E,\qquad
\mbox{where}\quad
h_k\equiv \max_{ l \in[l_0+k,l_0+k+1]}|h( l )|.
\end{equation*}
\end{teor}
\begin{proof}
Suppose the assertion is false. Then the series in the inequality converges for any choice of  $l_0$ and hence, first, $ h$  tends to zero at $l\to\infty$ and, second, the positive (recall that $h$ does not change its sign) function
\begin{equation*}
f( l ) \equiv  \frac{1}{h( l )}\int^{\infty}_ l  h(\breve l )\,\rmd \breve l  ,\qquad\quad l>0
\end{equation*}
is  defined. It is  bounded (by $E$) and the integral in its numerator  tends to zero at $l\to\infty$.
This enables us to use the Cauchy formula  \cite[n$^\circ$ 120]{Fiht1}  and find that
\[
f(l)= \frac{h(l_*) }{h'(l_*)  }\qquad\mbox{at some } l_*>l,
\]
By \eqref{eq: h'/h bound} this means that  $f$ is separated from zero
\begin{equation}\label{eq: f otd}
0< c_1 <f.
\end{equation}

It is easy to check that
$h'/h + f'/f=-1/f$
and hence condition \eqref{eq:ogr na f} is fulfilled.  Finally, the same equality   combined with the boundedness of $h'/h$, see \eqref{eq: h'/h bound}, and $1/f$, see \eqref{eq: f otd}, proves the boundedness of $f'/f$.
Thus all  conditions of lemma~\ref{lem: part of 8.5.5} hold, whence there must exist
a future inextendible  curve which, as  follows from \eqref{eq:Lam v comp},
is totally imprisoned in a  compact subset of $I^-(\gamma_0)\subset\ingh$. This again contradicts
\cite[proposition 6.4.7]{HawEl} because $\ingh$ is strongly causal.
\end{proof}

\end{document}